\documentstyle[12pt]{article}

\begin{document}
\title{Part of the D - dimensional anharmonic oscillator spectra}
\author{ Omar Mustafa and Maen Odeh\\
 Department of Physics, Eastern Mediterranean University\\
 G. Magusa, North Cyprus, Mersin 10 - Turkey\\
 email: omar.mustafa@mozart.emu.edu.tr\\
\date{}\\}
\maketitle
\begin{abstract}
{\small The pseudoperturbative shifted - $l$ expansion technique PSLET
[12,16] is generalized for states with arbitrary number of nodal zeros.
Interdimensional degeneracies, emerging from the isomorphism between
angular momentum and dimensionality of the central force Schr\"odinger
equation, are used to construct part of the $D$ - dimensional 
anharmonic oscillator bound - state spectra. PSLET results are found to
compare excellently with those from a series [7], exact and an open
perturbation [9] solutions.}
\end{abstract}
\newpage

\section{Introduction}

Besides its importance in quantum field theory and chemical physics, quantal
isotropic anharmonic oscillator\\
\begin{equation}
V(r)=\alpha_0 q^2+\alpha q^4,
\end{equation}\\
is also a popular theoretical laboratory for examining the validity of
approximation techniques [1-12]. Therefore, it has been a subject of an
enormous number of papers ( long list of these could be found in [3]).
However, most of these studies are devoted to one spatial dimension (1D).
Nevertheless, there has been a considerable interest, since the end of the
1960s as a result of studies in the far infrared and microwave regions, in
the analysis of two - and three - dimensional ( 2D and 3D, respectively)
anharmonic oscillators [6, and references therein]. Among the most recent
ones, to the best of our knowledge, exist the work of Lakshmanan et al.[6]
and Taseli [7]. They have treated the 2D - and 3D - anharmonic oscillators
via a phase - integral [6] (PIM) and a series solution, weighted by an
appropriate function, [7] (SSM) methods. Hence, further studies of such
oscillators should be of great interest.

In many problems, on the other hand, the Hamiltonian does not contain any
physical parameter suitable for a perturbation expansion treatment. More
often, the Hamiiltonian contains physical parameters, but, typically, zeroth
- order solutions for especial values of these are not tractable or good
starting approximations. Moreover, it is well known that the implementation
of Rayleigh - Schr\"odinger perturbation theory, or even naive perturbation
series, expresses the eigenvalue of the AHO as a formal power series in
$\alpha$ which is quite often divergent, or at best asymptotic, for every
$\alpha \neq 0$. One has therefore to resort to summation tools to sum up
such series [12-15]. Hence, apparently artificial perturbation theories have
been devised and proven to be ways to make progress [1,9,10,12,16-21, etc.].
Yet, in the simplest case, analytical calculations can aid numerical studies
in areas where numerical techniques might not be controlled. For example,
when bound - state wave functions with arbitrary nodal zeros are in demand
for certain singular potentials ( a next level of complexity), analytical
solutions can supply a basis for numerical calculations.

In a preceding paper [12], we have treated the nodeless states of the AHO
(1) via a pseudoperturbation shifted - $l$ expansion technique (PSLET),
where $l$ denotes the angular momentum quantum number. Successfully, the
same recipe has been applied to quasi - relativistic harmonic oscillator
 [16], and spiked harmonic oscillator, etc [17].

Encouraged by its satisfactory performance in handling nodeless states,
we feel tempted to
generalize PSLET recipe ( in section 2) for states with arbitrary number
of nodal zeros, $k \geq 0$. Moreover, in the underlying  time -
independent radial Schr\"odinger equation, in $\hbar=m=1$ units,\\
\begin{equation}
\left[-\frac{1}{2}\frac{d^{2}}{dq^{2}}+\frac{l(l+1)}{2q^{2}}+V(q)\right]
\Psi_{k,l}(q)=E_{k,l}\Psi_{k,l}(q),
\end{equation}\\
the isomorphism between orbital angular momentum $l$ and  dimensionality
$D$ invites interdimensional degeneracies [7,22-24]. Which, in
effect, allows us to generate the ladder of excited states for any given $k$
and nonzero $l$ from the $l$=0 result, with that $k$, by the transcription
$D \longrightarrow D+2l$. That is, if $E_{k,l}(D)$ is the eigenvalue in $D$
- dimensions, then\\
\begin{equation}
E_{k,l}(2) \equiv E_{k,l-1}(4) \equiv  \cdots 
\equiv E_{k,1}(2l) \equiv E_{k,0}(2l+2)
\end{equation}\\
for even $D$, and\\
\begin{equation}
E_{k,l}(3) \equiv E_{k,l-1}(5)  \equiv \cdots \equiv
E_{k,1}(2l+1) \equiv E_{k,0}(2l+3)
\end{equation}\\
for odd $D$. For more details the reader may refer to ref.s [7,22,25].
We therefore calculate, in section 3, the energies for 2D - and 3D -
anharmonic oscillators, for a given number of nodes $k$ and different
values of $l$, and construct part of its $D$ - dimensional bound - state
spectra. We compare our results with those from PIM [6], SSM [7], exact and
an open perturbation solutions [9]. Section 4 is devoted for concluding
remarks.

\section{The generalized PSLET}

Although some of the following expressions have appeared in previous
articles [12,16,17], we would like to repeat them to make this article self
contained.

We simply start with shifting the angular momentum quantum number $l$
in (2) through $\bar{l} = l - \beta$ and use $1/\bar{l}$ as a
pseudoperturbation expansion parameter. Where $\beta$ is a suitable shift
introduced to remove the poles that would emerge, at lowest orbital states
with $l=0$, in our expansions below. Hence, equation (2) reads\\
\begin{equation}
\left\{-\frac{1}{2}\frac{d^{2}}{dq^{2}}+\frac{\bar{l}^{2}+(2\beta+1)\bar{l}
+\beta(\beta+1)}{2q^{2}}+\frac{\bar{l}^2}{Q}V(q) \right\}
\Psi_{k,l} (q)=E_{k,l}\Psi_{k,l}(q),
\end{equation}\\
where Q is a constant that scales the potential $V(q)$ at large - $l_D$ limit
( the pseudoclassical limit [22]) and is set, for any specific choice of
$l_D$ and $k$, equal to $\bar{l}^2$ at the end of the calculations. Here
$l_D=l+(D-3)/2$, to incorporate the interdimensional degeneracies associated
with the isomorphism between angular momentum and dimensionality $D$. Hence,
$\bar{l} \longrightarrow \bar{l}=l_D - \beta$ through out this paper. Next,
we shift the origin of the coordinate system through
$x=\bar{l}^{1/2}(q-q_{o})/q_{o}$, where $q_o$ is currently an arbitrary
point to be determined below. Expansions about this point, $x=0$ (i.e.
$q=q_o$), localize the problem at  $q_o$ and the derivatives, in effect,
contain information not only at $q_o$ but also at any point on $q$-axis, in
accordance with Taylor's theorem.
Equation (5) thus becomes\\
\begin{equation}
\left[-\frac{1}{2}\frac{d^{2}}{dx^{2}}+\frac{q_{o}^{2}}{\bar{l}}
\tilde{V}(x(q))\right]
\Psi_{k,l}(x)=\frac{q_{o}^2}{\bar{l}}E_{k,l}\Psi_{k,l}(x),
\end{equation}\\
with\\
\begin{equation}
\frac{q_o^2}{\bar{l}}\tilde{V}(x(q))=
q_o^2\bar{l}\left[\frac{1}{2q_o^2}+\frac{V(q_o)}{Q}\right]
+\bar{l}^{1/2}B_1 x+\sum^{\infty}_{n=0} v^{(n)}(x) \bar{l}^{-n/2},
\end{equation}\\
where\\
\begin{equation}
v^{(0)}(x)=B_2 x^2 + \frac{2\beta+1}{2},
\end{equation}\\
\begin{equation}
v^{(1)}(x)=-(2\beta+1) x + B_3 x^3,
\end{equation}\\
\begin{eqnarray}
v^{(n)}(x)&=&B_{n+2}~ x^{n+2}+(-1)^n~ (2\beta+1)~ \frac{(n+1)}{2}~ x^n
\nonumber\\
&+&(-1)^{n}~ \frac{\beta(\beta+1)}{2}~ (n-1)~ x^{(n-2)}~~;~~n \geq 2,
\end{eqnarray}\\
\begin{equation}
B_n=(-1)^n \frac{(n+1)}{2}
+\left(\frac{d^n V(q_o)}{dq_o^n}\right)\frac{q_o^{n+2}}{n! Q}.
\end{equation}\\
It is then convenient to expand $E_{k,l}$
 as\\
\begin{equation}
E_{k,l}=\sum^{\infty}_{n=-2}E_{k,l}^{(n)}~\bar{l}^{-n}.
\end{equation}\\
Equation (6), along with (7)-(11), is evidently the one - dimensional
Schr\"odinger equation for a perturbed harmonic oscillator\\
\begin{equation}
\left[-\frac{1}{2}\frac{d^2}{dx^2}+\frac{1}{2}w^2 x^2 +\varepsilon_o
+P(x)\right]X_{k}(x)=\lambda_{k}X_{k}(x),
\end{equation}\\
where $w^2=2B_2$,\\
\begin{equation}
\varepsilon_o =\bar{l}\left[\frac{1}{2}+\frac{q_o^2 V(q_o)}{Q}\right]
+\frac{2\beta+1}{2}+\frac{\beta(\beta+1)}{2\bar{l}},
\end{equation}\\
and $P(x)$ represents the remaining terms in Eq.(6) as infinite power
series perturbations to the harmonic oscillator. One would then imply that\\
\begin{eqnarray}
\lambda_{k}&=&\bar{l}\left[\frac{1}{2}+\frac{q_o^2 V(q_o)}{Q}\right]
+\left[\frac{2\beta+1}{2}+(k+\frac{1}{2})w\right]\nonumber\\
&+&\frac{1}{\bar{l}}\left[\frac{\beta(\beta+1)}{2}+\lambda_{k}^{(0)}\right]
+\sum^{\infty}_{n=2}\lambda_{k}^{(n-1)}\bar{l}^{-n},
\end{eqnarray}\\
and\\
\begin{equation}
\lambda_{k} = q_o^2 \sum^{\infty}_{n=-2} E_{k,l}^{(n)}
\bar{l}^{-(n+1)}.
\end{equation}\\
Hence, equations (15) and (16) yield\\
\begin{equation}
E_{k,l}^{(-2)}=\frac{1}{2q_o^2}+\frac{V(q_o)}{Q}
\end{equation}\\
\begin{equation}
E_{k,l}^{(-1)}=\frac{1}{q_o^2}\left[\frac{2\beta+1}{2}
+(k +\frac{1}{2})w\right]
\end{equation}\\
\begin{equation}
E_{k,l}^{(0)}=\frac{1}{q_o^2}\left[ \frac{\beta(\beta+1)}{2}
+\lambda_{k}^{(0)}\right]
\end{equation}\\
\begin{equation}
E_{k,l}^{(n)}=\lambda_{k}^{(n)}/q_o^2  ~~;~~~~n \geq 1.
\end{equation}\\
Where $q_o$ is chosen to minimize $E_{k,l}^{(-2)}$, i. e.\\
\begin{equation}
\frac{dE_{k,l}^{(-2)}}{dq_o}=0~~~~
and~~~~\frac{d^2 E_{k,l}^{(-2)}}{dq_o^2}>0.
\end{equation}\\
Hereby, $V(q)$ is assumed to be well behaved so that $E_{k,l}^{(-2)}$ has
a minimum $q_o$ and there are well - defined bound - states.
Equation (21) in turn gives, with $\bar{l}=\sqrt{Q}$,\\
\begin{equation}
l_D-\beta=\sqrt{q_{o}^{3}V^{'}(q_{o})}.
\end{equation}\\
Consequently, the second term in Eq.(7) vanishes and the first term adds 
a constant to the energy eigenvalues. It should be noted that the energy term
$\bar{l}^2E_{k,l}^{(-2)}$  corresponds roughly to the energy of a classical
particle with angular momentum $L_z$=$\bar{l}$  executing circular motion of 
radius $q_o$ in the potential $V(q_o)$. It thus identifies the 
zeroth - order approximation, to all eigenvalues, as a classical 
approximation and the higher - order corrections as quantum fluctuations
around the minimum $q_o$, organized in inverse powers of $\bar{l}$.
The next correction to the energy series, $\bar{l}E_{k,l}^{(-1)}$,
consists of a constant term and the exact eigenvalues of the harmonic
oscillator $w^2x^2/2$.The shifting parameter
$\beta$ is determined by choosing
$\bar{l}E_{k,l}^{(-1)}$=0. This choice is physically motivated. In addition
to its vital role in removing the singularity at $l=0$, it also requires
the agreements between PSLET eigenvalues and eigenfunctions with
the exact well known ones
for the harmonic oscillator and Coulomb potentials.  Hence\\
\begin{equation}
\beta=-\left[\frac{1}{2}+(k+\frac{1}{2})w\right],~~
w=\sqrt{3+\frac{q_o V^{''}(q_o)}{V^{'}(q_o)}}
\end{equation}\\
where primes of $V(q_o)$ denote
derivatives with respect to $q_o$. Then equation (6) reduces to\\
\begin{equation}
\left[-\frac{1}{2}\frac{d^2}{dx^2} + \sum^{\infty}_{n=0} v^{(n)}
\bar{l}^{-n/2}\right]\Psi_{k,l} (x)= 
\left[\sum^{\infty}_{n=1} q_o^2 E_{k,l}^{(n-1)}
\bar{l}^{-n} \right] \Psi_{k,l}(x).
\end{equation}\\
Setting the wave functions with any number of nodes as \\
\begin{equation}
\Psi_{k,l}(x(q)) = F_{k,l}(x)~ exp(U_{k,l}(x)),
\end{equation}\\
equation (24) readily transforms into the following Riccati equation:\\
\begin{eqnarray}
&&F_{k,l}(x)\left[-\frac{1}{2}\left( U_{k,l}^{''}(x)+U_{k,l}^{'}(x)
U_{k,l}^{'}(x)\right)
+\sum^{\infty}_{n=0} v^{(n)}(x) \bar{l}^{-n/2} \right. \nonumber\\
&&\left. -\sum^{\infty}_{n=1} q_o^2 E_{k,l}^{(n-1)} \bar{l}^{-n} \right]
-F_{k,l}^{'}(x)U_{k,l}^{'}(x)-\frac{1}{2}F_{k,l}^{''}(x)=0,
\end{eqnarray}\\
where the primes denote derivatives with respect to $x$. It is
evident that this equation admits solution of the form \\
\begin{equation}
U_{k,l}^{'}(x)=\sum^{\infty}_{n=0} U_{k}^{(n)}(x)~~\bar{l}^{-n/2}
+\sum^{\infty}_{n=0} G_{k}^{(n)}(x)~~\bar{l}^{-(n+1)/2},
\end{equation}\\
\begin{equation}
F_{k,l}(x)=x^k +\sum^{\infty}_{n=0}\sum^{k-1}_{p=0}
a_{p,k}^{(n)}~~x^p~~\bar{l}^{-n/2},
\end{equation}\\
where\\
\begin{equation}
U_{k}^{(n)}(x)=\sum^{n+1}_{m=0} D_{m,n,k}~~x^{2m-1} ~~~~;~~~D_{0,n,k}=0,
\end{equation}\\
\begin{equation}
G_{k}^{(n)}(x)=\sum^{n+1}_{m=0} C_{m,n,k}~~x^{2m}.
\end{equation}\\
Substituting equations (27) - (30) into equation (26) implies\\
\begin{eqnarray}
&&F_{k,l}(x)\left[-\frac{1}{2}\sum^{\infty}_{n=0}\left(U_{k}^{(n)^{'}}
\bar{l}^{-n/2}
+ G_{k}^{(n)^{'}} \bar{l}^{-(n+1)/2}\right) \right. \nonumber\\
&-&\left.\frac{1}{2} \sum^{\infty}_{n=0} \sum^{n}_{m=0}
\left( U_{k}^{(m)}U_{k}^{(n-m)} \bar{l}^{-n/2}
+G_{k}^{(m)}G_{k}^{(n-m)} \bar{l}^{-(n+2)/2}
\right. \right.\nonumber\\
&+&\left.\left.2 U_{k}^{(m)}G_{k}^{(n-m)} \bar{l}^{-(n+1)/2}\right)
+\sum^{\infty}_{n=0}v^{(n)} \bar{l}^{-n/2}
-\sum^{\infty}_{n=1} q_o^2 E_{k,l}^{(n-1)} \bar{l}^{-n}\right] \nonumber\\
&-&F_{k,l}^{'}(x)\left[\sum^{\infty}_{n=0}\left(U_{k}^{(n)}\bar{l}^{-n/2}
+ G_{k}^{(n)} \bar{l}^{-(n+1)/2}\right)\right]-\frac{1}{2}F_{k,l}^{''}(x)
=0
\end{eqnarray}\\
The above procedure
obviously reduces to the one described by Mustafa and Odeh [12,16,17], when
$k=0$. Moreover, the solution of equation (31) follows from the uniqueness
of power series representation. Therefore, for a given $k$ we equate the
coefficients of the same powers of $\bar{l}$ and $x$, respectively. 
For example, when $k=1$ one obtains\\
\begin{equation}
D_{1,0,1}=-w,~~~ U_{1}^{(0)}(x) =-~w~x, 
\end{equation}\\
\begin{equation}
C_{1,0,1}=-\frac{B_{3}}{w},~~~~a_{0,1}^{(1)}=-\frac{C_{0,0,1}}{w},
\end{equation}\\
\begin{equation}
C_{0,0,1}=\frac{1}{w}\left(2C_{1,0,1}+2\beta+1\right),
\end{equation}\\
\begin{equation}
D_{2,2,1}=\frac{1}{w}\left(\frac{C_{1,0,1}^2}{2}-B_{4}\right),
\end{equation}\\
\begin{equation}
D_{1,2,1}=\frac{1}{w}\left(\frac{5}{2}~D_{2,2,1}+C_{0,0,1}~C_{1,0,1}
-\frac{3}{2}(2\beta+1)\right),
\end{equation}\\
\begin{equation}
E_{1,l}^{(0)} = \frac{1}{q_o^2}\left(\frac{\beta(\beta+1)}{2}+
a_{0,1}^{(1)}~C_{1,0,1}-\frac{3~D_{1,2,1}}{2}-\frac{C_{0,0,1}^2}{2}\right),
\end{equation}\\
etc. Here, we reported the nonzero coefficients only. One can then calculate
the energy eigenvalues and eigenfunctions from the knowledge of
$C_{m,n,k}$, $D_{m,n,k}$, and $a_{p,k}^{(n)}$ in a hierarchical manner.
Nevertheless, the procedure just described is suitable for a
software package such as  MAPLE to determine
the energy eigenvalue and eigenfunction corrections up to any order of the
pseudoperturbation series (12). 

Although the energy series, equation (12), could appear
divergent, or, at best, asymptotic for small $\bar{l}$, one can still 
calculate the eigenenergies to a very good accuracy by forming the 
sophisticated [N,M] Pad\'e approximation [1]\\
\begin{center}
$P_{N}^{M}(1/\bar{l})=(P_0+P_1/\bar{l}+\cdots+P_M/\bar{l}^M)/
(1+q_1/\bar{l}+\cdots+q_N/\bar{l}^N)$
\end{center}
to the energy series (12). The energy series (12) is calculated up to
$E_{k,l}^{(8)}/\bar{l}^8$ by
\begin{equation}
E_{k,l}=\bar{l}^{2}E_{k,l}^{(-2)}+E_{k,l}^{(0)}+\cdots
+E_{k,l}^{(8)}/\bar{l}^8+O(1/\bar{l}^{9}),
\end{equation}\\
and with the $P_{4}^{4}(1/\bar{l})$ Pad\'e approximant it becomes\\
\begin{equation}
E_{k,l}[4,4]=\bar{l}^{2}E_{k,l}^{(-2)}+P_{4}^{4}(1/\bar{l}).
\end{equation}\\
Our recipe is therefore well prescribed.

\section{D - anharmonic oscillator spectra}

In this section we consider the phenomenologically useful and methodically
challenging AHO interactions (1). and illustrate the above mentioned
procedure.

The substitution of (1) in (23), for $k \geq 0$, implies\\
\begin{equation}
w=\sqrt{\frac{8\alpha_o q_o+24\alpha q_o^3}{2\alpha_o q_o+4\alpha q_o^3}},
\end{equation}\\
and Eq.(22) yields\\
\begin{equation}
l_D+\frac{1}{2}\left(1+(2k+1)\sqrt{\frac{8\alpha_o q_o+24\alpha q_o^3}
{2\alpha_o q_o+4\alpha q_o^3}}\right)
=q_o^2 \sqrt{2\alpha_o+4\alpha q_o^2}.
\end{equation}\\
Once $q_o$ is determined (often numerically) the coefficients
$C_{m,n,k}$, $D_{m,n,k}$, and $a_{p,k}^{(n)}$ are obtained in a
sequential manner. Then, the
eigenvalues, equation (38), and eigenfunctions (25), along with (27)-(30),
are calculated in one batch for each value of $k$, $D$, $l$, $\alpha_o$,
and $\alpha$.

Tables 1 and 2 show the 2D - and 3D - AHO energies, respectively, for
different values of $k$, $l$, and $g=\alpha/2$. PSLET results, $E_P$,
compare excellently with those reported by Taseli [7] via a series
solution weighted by an appropriate function, $E_{SS}$. The 3D - AHO
energies for $k=1$ and $l=0$, or equivalently 1D - AHO third excited
state, are displayed along with those of Bessis and Bessis [9], $E_{BB}$,
via an open perturbation technique and the exact ones, $E_{ex}$, using
Bargman representation [26] ( direct numerical integrations) for different
anharmonicities. Obviously, our results are in quantitative and/or
qualitative agreements with the other ones. They are also in good agreement
with those reported by Kleinert [3] via a variational perturbation approach.

Clearly, the accuracy of PSLET increases for larger $k$ and/or $l$. The
$P_{4}^{4}(1/\bar{l})$ Pad\'{e} approximant enhances the accuracy, although
its effect is not dramatic for weak anharmonicities. Hence, one proceeds
with confidence and obtain, via (39), the 3D - energies ( table 4) for
states with $k=1$ and $l=1,5,10$. In doing so, one should keep the
stability of the Pad\'{e} squence in point ( for more details on this
issue the reader may refer to [16]). The same recipe is used to calculate
the 2D - AHO energies ( table 5) for states with $k=0,1$ and $l=0,1,5,10$.
Then the D - dimensional spectra follow from the implicated wisdom of
(3) and (4) (i.e.: the 2D - and 3D - AHO energies are the basic ingredients
for the construction of the D - dimensional ones [22]).

\section{Concluding remarks}

In this work we have generalized PSLET [12,16,17] for states with
arbitrary number of nodal zeros, $k \geq 0$. Starting with the "radical"
central force problem, represented by the radial Schr\"odinger equation, and
generalizing the angular momentum to the D - Dimensional one
( i.e.: $l \longrightarrow l_D=l+(D-3)/2$), we have treated the AHO in
D - dimensions. The comparison between PSLET results with the other ones,
including direct numerical integration, is readily satisfactory.

Although we have used Pad\'{e} approximants to improve the numerical
performance of PSLET, it is by no means clear whether the Pad\'{e}
approximants are necessarily the most effective ones. It has been suggested
by Weniger [14] that much better results could be obtained via a class of
sequence transformations [27] ( The details of which could be found in
[27-30, and references therein]). However, such interesting
investigations already lie beyond the scope of the attendant proposal.

Finally, the applicability of PSLET extends far beyond the present
D - dimensional anharmonic oscillator model. It could be applied to
angular momentum states of multi - electron atoms [23,24], quark - antiquark
models [31], 2D - Hydrogenic donor states in an arbitrary magnetic fields
[32], 2D - magnetoexcitons [33], two - electron quantum dots [34], etc.

\newpage
\begin{table}
\begin{center}
\caption{ 2D anharmonic oscillator (1) energies, in $\hbar=m=1$ units.
Where $\alpha_0=1/2$ and $g=\alpha/2$ to recover Taseli's results,
$E_{SSM}$, via series solution [7]. $E_P$ represents PSLET results, Eq.(38),
and $E[4,4]$ shows the $P_{4}^{4}(1/\bar{l})$ Pad\'{e} approximant,Eq.(39).}
\vspace{1cm}
\begin{tabular}{|ccclll|}
\hline\hline
 $g$ & $k$ & $l$ & 2$E_P$ & 2$E[4,4]$ & $E_{SSM}$\\
\hline
 $10^{-4}$ & 0 & 0 & 2.000199955022 & 2.000199955022 & 2.000199955022\\
           &   & 5 & 12.00419695953 & 12.00419695953 & 12.00419695953\\
           & 1 & 1 & 8.002398591662 & 8.002398591662 & 8.002398591662\\
           & 3 &   & 16.00958904459 & 16.00958904459 & 16.00958904459\\
 1         & 0 & 0 & 2.947835       & 2.952052       & 2.952050 \\
           &   & 2 & 10.390626203   & 10.390627276   & 10.390627295\\
           &   & 4 & 19.217523488   & 19.2175234955  & 19.2175234959\\
           & 1 & 1 & 15.48277174    & 15.48277148    & 15.48277158\\
 $10^{4}$  & 0 & 0 & 50.75164       & 50.54788       & 50.54804\\
           &   & 4 & 368.030083     & 368.030082436  & 368.030082448\\
           & 1 & 0 & 205.3783       & 205.3774       & 205.3777\\
           &   & 2 & 394.577414     & 394.577403     & 394.577407\\
\hline\hline
\end{tabular}
\end{center}
\end{table}
\newpage
\begin{table}
\begin{center}
\caption{ Same as Table 1 for 3D anharmonic oscillator (1) energies.}
\vspace{1cm}
\begin{tabular}{|cccccc|}
\hline\hline
 $g$ & $k$ & $l$ & 2$E_P$ & 2$E[4,4]$ & $E_{SSM}$\\
\hline
 $10^{-4}$ & 0 & 0  & 3.0003748969 & 3.0003748969 & 3.0003748969 \\
           &   & 10 & 23.014356719 & 23.014356719 & 23.014356719 \\
 1         &   & 0  & 4.648511     & 4.648815     & 4.648813 \\
           &   & 1  & 8.380337     & 8.38034245   & 8.38034253 \\
           &   & 5  & 26.528917558 & 26.528917558 & 26.528917558 \\
           & 1 & 3  & 27.898417763 & 27.898417756 & 27.898417760\\
 20        & 0 & 1  & 19.783266    & 19.7832518   & 19.7832519\\
           &   & 2  & 30.057200    & 30.057199029 & 30.057199045\\
           &   & 5  & 65.961500037 & 65.96150003049 & 65.96150003068\\
           & 1 & 1  & 44.209282    & 44.209279007   & 44.209279973\\
 $10^{3}$  & 0 & 0  & 38.092       & 38.086822      & 38.086833\\
           & 0 & 3  & 149.439046   & 149.439045568  & 149.439045581\\
\hline\hline
\end{tabular}
\end{center}
\end{table}

\newpage
\begin{table}
\begin{center}
\caption{ Three-dimensional energies for the $k=1$ and $l=0$ state or
equivalently one-dimensional third excited state energies for 
$V(q)=\frac{q^{2}}{2}+\alpha q^{4}$. $E_{BB}$ denotes Bessis and Bessis
results [9] and the exact ones $E_{exact}$, reported therein, from direct
numerical integrations for different anharmonicities.}
\vspace{1cm}
\begin{tabular}{|ccccc|}
\hline\hline
 $\alpha$ & $E_P$ & $E[4,4]$ & $E_{ex}$ & $E_{BB}$ \\
\hline
 0.002    &3.53674413 & 3.536744133& 3.53674413 &3.53674\\
 0.01     &3.67109494 & 3.67109494 & 3.67109494 &3.67109\\
 0.1      &4.6288828  & 4.6288828  & 4.62888281 &4.62884\\
 0.3      &5.79657376 & 5.79657363 & 5.79657363 &5.79679\\
 0.5      &6.578402   & 6.578402   & 6.57840195 &6.57953\\ 
 0.7      &7.193266   & 7.193265   & 7.19326528 &7.19549\\
 1        &7.942405   & 7.942404   & 7.94240399 &7.94630\\
 2        &9.727325   & 9.727322   & 9.72732319 &9.73596\\
 50       &27.192660  & 27.192638  & 27.1926458 &27.2473\\
 1000     &73.419158  & 73.419089  & 73.419114  &73.5805\\
 8000     &146.745600 & 146.745461 & 146.745512 &147.0714\\
\hline\hline
\end{tabular}
\end{center}
\end{table}

\newpage
\begin{table}
\begin{center}
\caption{ Three-dimensional energies for states with $k=1$ and $l=1,5,10$ 
for $V(q)=\frac{q^{2}}{2}+\alpha q^{4}$. Only $E[4,4]$'s are listed for
different anharmonicities.}
\vspace{1cm}
\begin{tabular}{|cccc|}
\hline\hline
$\alpha$  &  $l=1$        & $l=5$          & $l=10$        \\
\hline
0.01    & 4.76645813712 & 9.289594583372 & 15.233049583486\\
0.1     & 6.176138      & 12.89579856    & 22.309686916    \\
0.5     & 8.93090       & 19.3542918     & 34.30436531     \\
1       & 10.83313      & 23.7006578     & 42.25455311     \\
50      & 37.4108       & 83.258353      & 149.9642236 \\
1000    & 101.07403     & 225.231013     & 405.9901767\\
\hline\hline
\end{tabular}
\end{center}
\end{table}

\newpage
\begin{table}
\begin{center}
\caption{ Two-dimensional energies for states with $k=0,1$ and $l=0,1,5,10$ 
for $V(q)=\frac{q^{2}}{2}+\alpha q^{4}$. Only $E_[4,4]$'s are listed for
different anharmonicities.}
\vspace{1cm}
\begin{tabular}{|cccccc|}
\hline
$k$ & $\alpha$  & $l=0$        &$l=1$        &$l=5$        &$l=10$  \\
\hline
 0 & 0.01 & 1.0191783021 & 2.056555600 & 6.372257220 & 12.0962676139707\\
   & 0.1  & 1.150188     & 2.4143403   & 8.29606606  & 16.976887733   \\
   & 0.3  & 1.33966      & 2.895905    & 10.53678440 & 22.227347003    \\
   & 0.5  & 1.4760       & 3.231453    & 12.01658310 & 25.611647809     \\
   & 0.7  & 1.5866       & 3.499749    & 13.17451055 & 28.235574336     \\
   & 1    & 1.7242       & 3.830324    & 14.58077151 & 31.403160969   \\
   & 50   & 5.512        & 12.6399     & 50.376652   & 110.564887242   \\
\hline
 1 & 0.01 & 3.129048426  & 4.21691935465& 8.71236579303& 14.62528496652399\\
   & 0.1  & 3.876642     & 5.3954269    & 12.01280444  & 21.323943303\\
   & 0.3  & 4.8105       & 6.80426      & 15.6275228   & 28.28370297\\
   & 0.5  & 5.4412       & 7.74139      & 17.9691075   & 32.72664974\\
   & 0.7  & 5.9389       & 8.47655      & 19.7879425   & 36.15909940\\
   & 1    & 6.5466       & 9.3708       & 21.9862477   & 40.29318318\\
   & 50   & 22.267       & 32.237       & 77.122811    & 142.8926171\\
\hline\hline
\end{tabular}
\end{center}
\end{table}

\end{document}